\documentclass[%
reprint,
superscriptaddress,
 amsmath,amssymb,
 aps,
 two column
]{revtex4-2}

\usepackage{graphicx}
\usepackage{bm, physics}
\usepackage{float}
\usepackage{mathtools}
\usepackage{hyperref}

\usepackage{color}

\let\svthefootnote\thefootnote
\newcommand\freefootnote[1]{%
  \let\thefootnote\relax%
  \footnotetext{#1}%
  \let\thefootnote\svthefootnote%
}

\begin{document}

\title{Inverse magneto-conductance design by automatic differentiation}
\author{Yuta Hirasaki}
\email{yutah2@illinois.edu}
 \affiliation{ 
Department of Applied Physics, The University of Tokyo, Tokyo 113-8656, Japan.
}%
 \affiliation{ 
Department of Physics, University of Illinois at Urbana-Champaign, Urbana, Illinois 61801, USA.
}%
\author{Koji Inui}
\email{koji-inui@issp.u-tokyo.ac.jp}
\thanks{\\ Y. H. and K. I. contributed equally to this work.}
 \affiliation{ 
Department of Applied Physics, The University of Tokyo, Tokyo 113-8656, Japan.
}%
\affiliation{Department of Nuclear Engineering and Management, The University of Tokyo, Tokyo 113-8656, Japan.}
\affiliation{Institute for Solid State Physics, The University of Tokyo, Kashiwa 277-8581, Japan.}

\author{Eiji Saitoh}
\affiliation{ 
Department of Applied Physics, The University of Tokyo, Tokyo 113-8656, Japan.
}%
\affiliation{Institute for AI and Beyond, The University of Tokyo, Tokyo 113-8656, Japan.}
\affiliation{WPI Advanced Institute for Materials Research, Tohoku University, Sendai 980-8577, Japan.}
\affiliation{RIKEN Center for Emergent Matter Science (CEMS), Wako 351-0198, Japan.}

\date{\today}

\begin{abstract}
Magneto-conductance in thin wires often exhibits complicated patterns due to the quantum interference of conduction electrons. These patterns reflect microscopic structures in the wires, such as defects or potential distributions. In this study, we propose an inverse design method to automatically generate a microscopic structure that exhibits desired magneto-conductance patterns. We numerically demonstrate that our method accurately generates defect positions in wires and can be effectively applied to various complicated patterns. We also discuss techniques for designing structures that facilitate experimental investigation.
\end{abstract}


\maketitle 

\section{Introduction}
\begin{figure*}[t]
\includegraphics{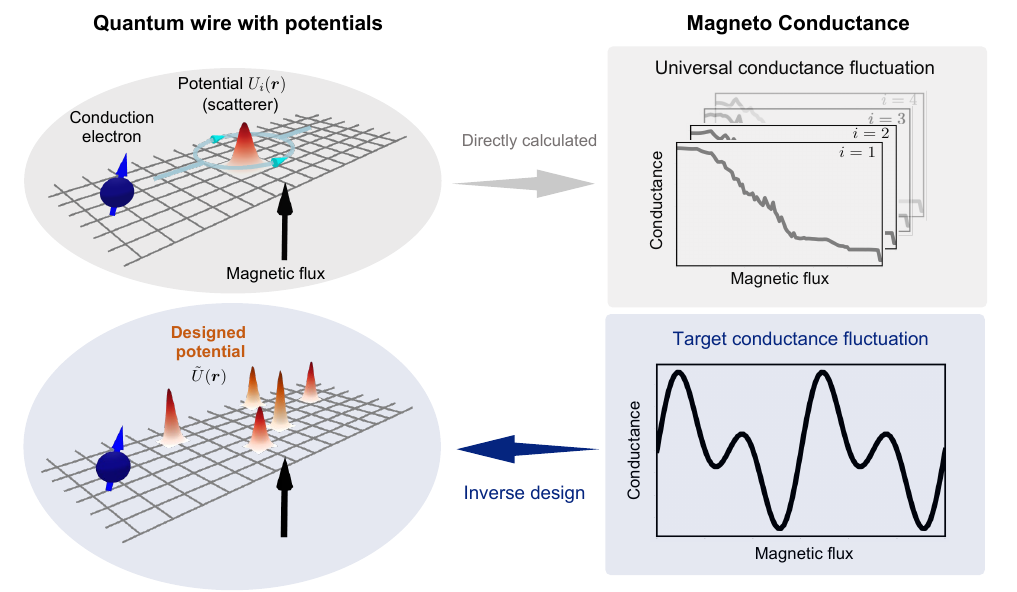}
\caption{Schematic picture of inverse design of magneto-conductance fluctuation. The top panels (colored gray) illustrate a conventional procedure; we calculate a magneto-conductance fluctuation from a given potential distribution $U_i(\bm{r})$, where $i = 1, 2, \dots$ is the label of the potential distributions. For each potential distribution, the different patterns of magneto-conductance fluctuation are obtained, as shown in the upper right panel. The bottom panels illustrate our inverse design framework. We determine a designed onsite potential $\tilde{U}(\bm{r})$ based on a given target conductance pattern.
}
\label{f1}
\end{figure*}

Quantum wires are quasi-one-dimensional structures whose electrical conductance is heavily influenced by quantum interference. A fundamental example is the Aharonov-Bohm (AB) effect \cite{PhysRev.115.485, PhysRevLett.54.2696}, where an electromagnetic potential influences the phase of the electron wave function, and the interference between different electron paths modulates the conductance. In real quantum wires, scatterers such as impurities act in a similar way to holes in the AB effect; the application of the external magnetic fields to the quantum wires leads to complex magneto-conductance fluctuation, known as universal conductance fluctuation (UCF), due to the modulation of the phases of the wave functions\cite{RevModPhys.57.287}. The upper panel of Fig.~\ref{f1} schematically illustrates the UCF patterns varying with the potential distribution $U_i(\bm{r})$, representing the scatterers in the quantum wires, where $i$ is the label of the respective distributions. 

The UCF patterns reflect the microscopic structures, such as the shape of the wire, impurities, and potential distributions, through the modulation of the wave functions\cite{PhysRevB.96.134201}; these patterns are called quantum fingerprints since they contain information about these microscopic structures. In principle, one can infer the microscopic structure from a given UCF pattern by solving inverse problems. However, in most cases, the UCF patterns are so complicated that it is difficult to predict the microscopic structure of quantum wires. 

Several works have addressed this inverse problem by employing machine learning (ML) methods \cite{PhysRevB.89.235411, PhysRevB.102.064205, SDncom}. Although potentially powerful, ML methods require large amounts of data to build models with good generalization performance; this approach is limited by the difficulty of collecting data on the corresponding microstructures and UCFs.

An alternative approach is to leverage automatic differentiation, which was recently developed by one of the authors\cite{inui2023inverse, inui2024inverse}, to design Hamiltonians that reproduce target properties. {Automatic differentiation is a computational technique that computes the derivatives of a wide range of functions with analytical accuracy by systematically tracing each numerical operation. This method has been applied in neural network optimization, most notably in the backpropagation algorithm, by leveraging the gradient descent method.} Automatic differentiation allows us to solve the inverse problem without preparing training data, or relying on black box models such as neural networks. In addition, its computational cost remains manageable in practice, even when the number of parameters increases to, for example, a million. Previous studies have successfully designed systems with large quantum entanglement \cite{inui2024inverse}, a large anomalous Hall effect, a photovoltaic effect \cite{inui2023inverse}, {and a non-equilibrium Green function} \cite{PhysRevB.108.195143}. Therefore, it can be used to identify microstructure that reproduces the target UCF. However, this approach has been applied primarily only to the diagonalization of the Hamiltonians of isolated systems. Application to dynamical properties in more general systems was left unexplored.

In this study, we propose an inverse design framework that leverages automatic differentiation to identify a microscopic structure from a given magneto-conductance fluctuation. Specifically, this framework determines the appropriate onsite potentials, representing a microscopic structure, that realizes a given magneto-conductance pattern, as illustrated in the lower panel of Fig. \ref{f1}. Optimization using automatic differentiation can handle a large number of parameters, allowing for flexible inverse design even if the target pattern is very complicated. This framework is general and applicable for purposes other than magneto-conductance patterns. It can also be used to optimize Hamiltonian parameters other than onsite potentials. In addition, although inverse problems are generally ill-defined and thus have multiple possible solutions, we have developed an additional method to derive a more experimentally feasible solution among them. Therefore, this framework is not only capable of solving complex inverse problems but can lead to the development of systems with the desired properties.

This paper is organized as follows. In Sec. \ref{method}, we introduce the framework. We detail its setup, the conductance calculation, and the whole flow of the framework. Sec. \ref{Res} is devoted to results. In Sec. \ref{ResA}, we demonstrate that our framework can automatically determine the position of a defect in a quantum wire from a given UCF as a proof of concept. In Sec. \ref{ResB}, we show that our framework can be applied to arbitrarily complex patterns, using photographs of Nirvana and the Colosseum as the subjects. In Sec. \ref{ResC}, we perform additional optimizations using a different objective function to transform the optimized structure into a more experimentally friendly structure without changing the magneto-electric patterns. Finally, we give the concluding remarks in Sec. \ref{Summary}.

\section{Method}\label{method}
In this section, we introduce a method for constructing a microscopic structure in a quantum wire that realizes a given magneto-conductance fluctuation based on the inverse Hamiltonian design framework. 
We combine the Python transport package KWANT \cite{groth2014kwant} with the Python automatic differentiation package JAX \cite{bradbury2018jax}. 
The KWANT allows us to calculate the magnetic field dependence of the conductance of quantum wires. We have rewritten parts of KWANT using JAX to introduce automatic differentiation capabilities.

\subsection{Magneto-conductance calculation}
In this study, quantum wires are modeled by a tight-binding model with a two-dimensional scattering region, and semi-infinite leads are connected at both ends. {Note that each site in the tight-binding model can be interpreted as a finite-difference discretization of a continuum Hamiltonian, where the model operates at a scale beyond the atomic level \cite{groth2014kwant}.} The effect of the magnetic field is taken into account as a Peierls phase with the Landau gauge. The Hamiltonian is given by
\begin{align}\label{eq1}
\begin{split}
\hat{H} &= \sum_{i}h_i c_i^\dagger c_i \\
&- \sum_{\langle i, j \rangle}t\exp\left[-i\pi \frac{Ba^2}{\phi_0}(x_i - x_j)(y_i + y_j)\right]c_i^\dagger c_j,
\end{split}
\end{align}
where $c_i$ ($c_i^\dagger$) is an annihilation (creation) operator for an electron at site $i$. $x_i, y_i$ and $h_i$ denote $x$ axis, $y$ axis, and onsite potential at site $i$, respectively. $t$ is a hopping energy, $a$ is a lattice constant, and $\phi_0 \coloneqq \frac{h}{e}$ is the magnetic flux quantum. We apply the magnetic flux $B$ to the entire system including a scattering region and leads. KWANT calculates the conductance of a given tight-binding Hamiltonian using the Landauer–Büttiker formula \cite{datta1997electronic}. 
The conductance is given by,
\begin{align}
    G =\frac{e^2}{h}\sum_{n\in a, m\in b}|S_{nm}|^2, 
\end{align}
where $S_{nm}$ is the scattering matrix, and $a$ ($b$) denote an input (output) lead. See Ref. \cite{groth2014kwant} for the details.

We use a tight-binding model on a square lattice. Throughout the paper, we use the following settings. The size of the scattering region is $30\times 30$ sites and two leads are attached to both ends of the scattering region with its width to be $30$ sites. The lattice constant and the hopping energy are set to $a = 1$ and $t = 1$, respectively. The electron energy (the parameter in the r.h.s. of Eq. (6) in Ref. \cite{groth2014kwant}) is fixed at $E = 1.2$. We vary the magnetic field so that $\pi\frac{Ba^2}{\phi_0}$ is swept from $0$ to $0.12$ in 101 divisions. The parameters to be optimized are the onsite potentials $h_i$ at the scattering region; each site can have different values. The onsite potentials at the leads are fixed at $h_i = 4t = 4$. 

\subsection{Calculation flow and computational detail}\label{methodOpt}
\begin{figure}
  \includegraphics{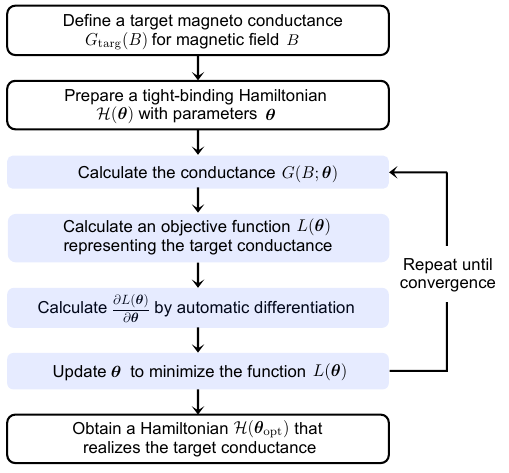}
  \caption{Flowchart of the inverse design of a magneto-conductance fluctuation using automatic differentiation. See the main text for details.}
  \label{f2}
\end{figure}

The overall optimization flow is illustrated in Fig. \ref{f2}. First, we define a target magneto-conductance $G_\mathrm{targ}(B)$. We then prepare a tight-binding Hamiltonian with parameters $\bm{\theta}$. Once the Hamiltonian is prepared, the conductance $G(B; \bm{\theta})$ is directly calculated using KWANT combined with JAX. Next, we define the objective function $L(\bm{\theta})$ that represents the desired properties. $L(\bm{\theta})$ basically corresponds to the difference between $G(B; \bm{\theta})$ and $G_\mathrm{targ}(B)$. The details of $L(\bm{\theta})$ are provided in Sec. \ref{Res}. After the objective function is computed, we calculate the partial derivatives $\frac{\partial L}{\partial \bm{\theta}}$ using automatic differentiation. Then we update the parameters $\bm{\theta}$ using a gradient descent method to minimize $L(\bm{\theta})$. We iteratively apply this procedure until the objective function converges. With the successfully optimized parameters $\bm{\theta}_\mathrm{opt}$, the Hamiltonian $H(\bm{\theta}_\mathrm{opt})$ should reproduce the target magneto-conductance.

For the gradient descent method, we employ the RMSProp optimizer \cite{reddy2018optimization} in Sec.\ref{ResA} and the AdaBelief optimizer \cite{NEURIPS2020_d9d4f495} in Sec. \ref{ResB} and \ref{ResC}. We adjust the learning rates $\eta$ during optimization processes. In Sec.\ref{ResA}, we start with $\eta = 10^{-2}$ and decrease it as it follows; at $2000$ steps, we multiply $\eta$ by $0.9$, at $5000$ steps by $0.5$, at $10000$ by $0.1$, and at $15000$ by $0.1$. In Sec. \ref{ResB} and \ref{ResC}, we start with $\eta = 10^{-2}$ and decrease it in the following way; at 2,000 steps, $\eta$ is multiplied by 0.9, at 5,000 by 0.5, at 10,000 by 0.1, at 15,000 by 0.1, and at 30,000 by 0.1. We use the default values set in \texttt{jax.example\_libraries.optimizers} regarding the other hyperparameters. All computations are executed on an NVIDIA A100 GPU.

\section{Results}\label{Res}
\subsection{Identification of Defect}\label{ResA}
\begin{figure*}
  \includegraphics{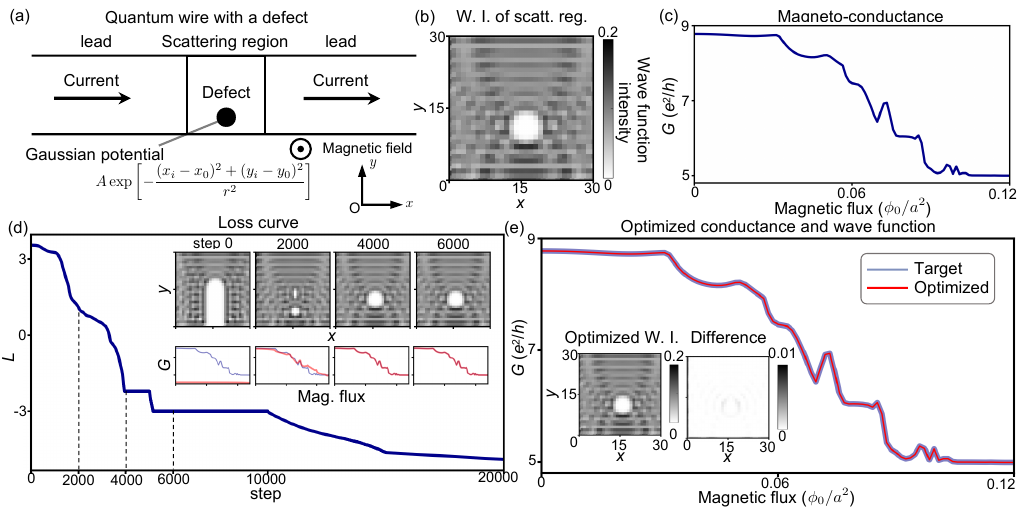}
  \caption{Inverse problem solving of a defect position from a measured conductance fluctuation. (a) A schematic picture of the system. The sample extends infinitely in the $x$ direction. We apply a uniform magnetic flux perpendicularly to the sample. There is a defect in the sample represented by a Gaussian potential. (b) The WI of a scattering region in the sample. (c) The magneto-conductance of the sample as a function of the magnetic flux. (d) The objective function in the optimization process. (e) The magneto-conductance after the optimization. 
  }
  \label{f3}
\end{figure*}

First, we validate our framework by applying it to a straightforward task: detecting the position of a defect in a quantum wire, as illustrated in Fig. \ref{f3}. Initially, we generate the UCF pattern of a quantum wire with a single defect using KWANT. We then assess the accuracy of our framework in identifying the defect's position from the UCF pattern.

The system is schematically shown in Fig. \ref{f3}(a). Here, an antidot defect in a quantum wire is modeled by a Gaussian-type onsite potential distribution, where electrons are scarcely present in regions with high potential. The onsite potentials on the scattering region are defined by 
\begin{align}
    h_i = 4t + A\exp\left(-\frac{(x_i-x_0)^2 + (y_i-y_0)^2}{r^2}\right), 
\end{align}
where $A$, $(x_0, y_0)$ and $r$ represent the amplitude, center position, and radius of the Gaussian distribution, respectively. The parameters are set as follows: $A = 10$, $x_0 = 15$, $y_0 = 10$, and $r = 2$.

Figure \ref{f3}(b) shows the wave-function intensity (WI) of the scattering region with the magnetic field $B = 0$. Each pixel represents the local density of state at each site. It is evident that the WI is nearly zero in the region of the Gaussian potential. The target magneto-conductance $G_\mathrm{targ}(B)$ is shown in Fig. \ref{f3}(c).

We then apply the inverse design framework to reconstruct the Gaussian potential from the target magneto-conductance $G_\mathrm{targ}(B)$. The objective function is defined as
\begin{align}\label{eq2}
    L(\bm{\theta}) = \log\left(\sqrt{\sum_{B}[G(B; \bm{\theta}) - G_\mathrm{targ}(B)]^2}\right),
\end{align}
which represents the difference between $G(B, \bm{\theta})$ and $G_{\mathrm{targ}}(B)$. The logarithm function is employed to prevent the objective function from becoming too small, thereby avoiding the vanishing gradient problem.

When solving the inverse problem, the choice of parametrizations plays a crucial role in determining the success of the optimization. A straightforward approach is to use a single Gaussian-type potential, with $\bm{\theta} =(A, x_0, y_0, r)$ as the optimization parameters. However, this approach was unsuccessful in identifying the defect, as it became trapped in local minima during the optimization process. To address this issue, we introduce multiple Gaussian potentials in the region and optimize all parameters simultaneously. The onsite potential is parametrized as 
\begin{align}
    h_i(\bm{\theta}) = 4t + \sum_{k=1}^n A_k \exp\left(-\frac{(x_i - x_k)^2 + (y_i - y_k)^2}{r_k^2}\right);
\end{align}
with the optimization parameter set $\bm{\theta} = (y_1, y_2, \dots, y_n, A_1, A_2, \dots, A_n, r_1, r_2, \dots, r_n) \in \mathbb{R}^{3n}$. We set $n = 10$, and the initial $y$-coordinates $(y_1, y_2, \dots, y_n)$ are equally spaced from $y_1 = 0$ to $y_n = 15$, with the initial amplitudes and radii set as $A_k = 1$ and $r_k = 2$ for all $k = 1, 2, \dots, n$, respectively. Note that shifting the onsite potential in the $x$ direction does not change the result due to translational symmetry, and thus we set the $x$ coordinates $x_k$ to be constant for all $k = 1, 2, \dots, n$. In addition, due to the mirror symmetry with respect to the $y$-axis, only half of the region in the $y$-component needs to be considered; we restrict the $y$-coordinates to be below $y=15$. 

The optimization process of the objective function $L(\bm{\theta})$ is shown in Fig. \ref{f3}(d). The value of $L$ decreases monotonously from above $3$ to below $-3$, indicating the optimization is successfully performed. The inset shows the WIs of the scattering region (top) and the corresponding magneto-conductance pattern (bottom) at optimization steps $0, 2000, 4000, 6000$. The red (blue) lines in the bottom panels show the calculated (target) magneto-conductance pattern. As the optimization progresses, the calculated pattern increasingly approaches the target pattern. Figure \ref{f3}(e) shows the magneto-conductance pattern after the optimization. The optimized magneto-conductance pattern (red) is almost identical to the target pattern (blue). The left panel in the inset shows the WI of the scattering region with the optimized parameters. The WI is also almost identical to the WI of the target shown in Fig. \ref{f3}(b). To clearly show that, the difference in WI is plotted to the right of the inset of Fig. \ref{f3}(e). See the supplementary material for the optimization process and the optimized potential distribution in detail.

\subsection{Complicated Conductance Patterns}\label{ResB}
\begin{figure}
\includegraphics{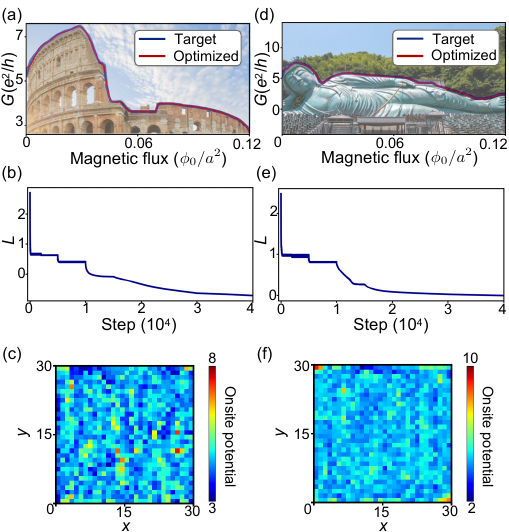}
\caption{
Reproduction of complicated magneto-conductance patterns by optimizaing onsite potentials. (a)(d) Target and optimized conductance patterns. The target patterns are formed by tracing the shapes of Colosseum (a) and nirvana (d). (b)(e) The optimization process of the objective functions. (c)(f) The values of the optimized onsite potentials. 
}
\label{f4}
\end{figure}

Next, we demonstrate that our framework can be applied to arbitrarily complex magneto-conductance patterns. As target magneto-conductance patterns, we have prepared two artificial patterns by tracing the shape of a Colosseum and Nirvana (a lying big Buddha), as indicated by the blue lines in Fig. \ref{f4}(a) and (d). To allow for as flexible optimization as possible, the onsite potentials for each site, $h_i$, in the scattering region are treated as an independent parameter as $\bm{\theta} = (h_1, h_2, \dots, h_{900}) \in \mathbb{R}^{900}$. The objective function $L(\bm{\theta})$ is the same as Eq. \ref{eq2}. {The optimization of the onsite potentials corresponds to adjusting the gate voltages applied to the sample.}

Figure \ref{f4}(b) and (e) show the optimization process of the objective function for the Colosseum and Nirvana, respectively. In both cases, the values of the objective function decrease monotonously. The optimized magneto-conductance patterns are plotted by the red lines in Fig. \ref{f4}(a) and (d). They are in good agreement with the target patterns. The optimized onsite potentials are shown in Fig. \ref{f4}(c) and (f). 

Our results show that the framework can successfully reproduce the target patterns by optimizing the potential distributions. This highlights the versatility of our approach in handling various types of microscopic structures. We note that the solutions are not unique; during the numerical calculation, we have found several onsite potential distributions that reproduce the same magneto-conductance pattern by starting with different initial parameters. {Different initial conditions yield different optimized parameters, but all of them reproduce the target conductance well.}  We also note that it is rather challenging to reproduce a pattern in which the conductivity increases when the magnetic field is increased because, in general, the conductivity decreases as the magnetic field due to the localization of electrons.

\subsection{Refining structure design}\label{ResC}

\begin{figure*}
\includegraphics{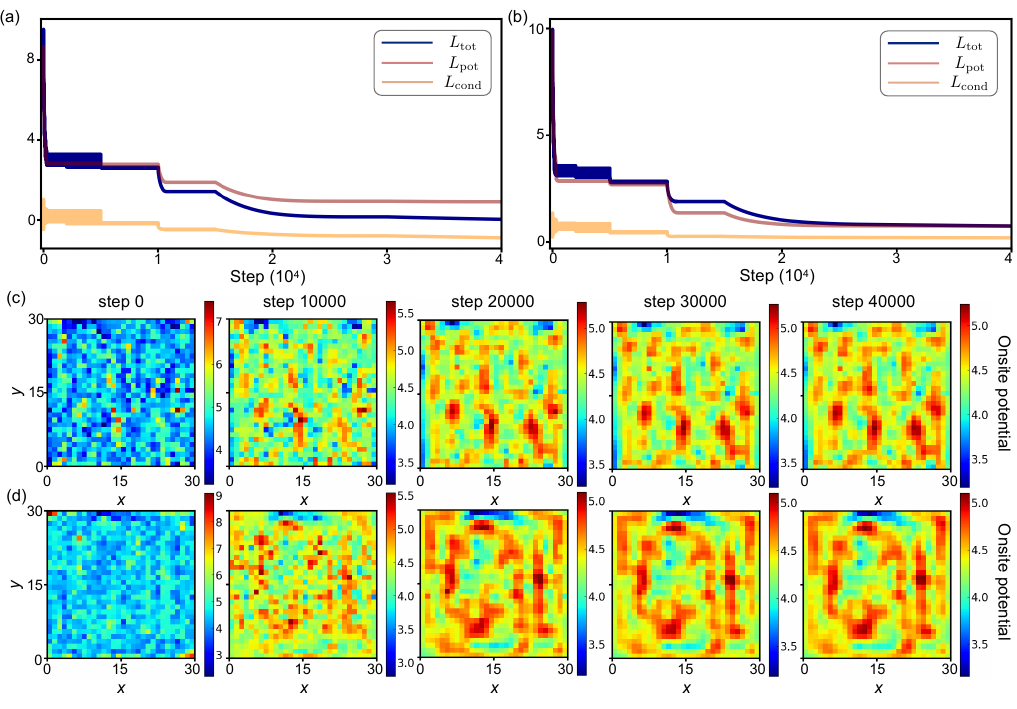}
\caption{Adjustment of the onsite potential by adding a regularization term. The objective function in the optimization process for the target conductance of (a) Colosseum and (b) Nirvana. The changes in the onsite potentials for (c) Colosseum and (d) Nirvana are depicted at each 10,000 step. 
}
\label{f5}
\end{figure*}
In this section, we discuss a technique to adjust the potential distributions to a more coarse-grained structure. Straightforward application of our framework tends to yield fine and irregular potential distributions, as shown in Fig. \ref{f4}(c) and (f); these patterns are difficult to design in experiments. Here, we demonstrate that we can control the complexity of the structure by incorporating a regularization term into the objective function. 

The regularization term $L_\mathrm{pot}(\bm{\theta})$ is given by  
\begin{align}
    L_\mathrm{pot}(\bm{\theta}) = \sum_{<i, j>}(h_{i} - h_{j})^2,
\end{align}
where the summation is taken over neighboring sites. This term favors a smoother potential distribution as the values of neighboring potentials approach each other. 

The total objective function is given by
\begin{align}\label{eq_tot}
    L_\mathrm{tot} = L_\mathrm{cond}(\bm{\theta}) + \gamma L_\mathrm{pot}(\bm{\theta}),
\end{align}
where $L_\mathrm{cond}$ is the same as Eq. \ref{eq2} and $\gamma$ is a hyperparameter. We set $\gamma = 10^{-2}$. We perform the optimization using Eq. \ref{eq_tot} after the optimization using Eq. \ref{eq2}. 

We apply this technique to the optimized results of the Colloseum and Nirvana shown in Sec. \ref{ResB}. The results are shown in Fig. \ref{f5}. In Fig. \ref{f5}(a) [(b)], the optimization process of the objective function for Colosseum [Nirvana] is shown. The blue line shows the total objective function $L_\mathrm{tot}$, and the red (orange) line shows $L_\mathrm{pot}$ ($L_\mathrm{cond}$). In both cases, $L_\mathrm{pot}$ decreases significantly, while $L_\mathrm{cond}$ also shows a slight decline in value. These suggest that the potential distributions get smoother while the magneto-conductance patterns are maintained. The changes in the onsite potentials at every 10,000 steps are depicted in Figs. \ref{f5}(c) and (d). As the optimization proceeds, the onsite potentials become more continuous and less inhomogeneous (also see the changes in the scale of the color bars). 

We have demonstrated that we can adjust the onsite potential distribution by adding a regularization term. {When controlling the onsite potential using an electric field, a coarser structure, such as that on the right side of Fig. \ref{f5}, is more feasible to realize than the finer structure on the left. Thus, this technique enables us to identify the on-site potential distributions that are easier to achieve among multiple solutions that reproduce magneto-conductance fluctuation.} Further sophistication in the regularization term would allow for the design of more experimentally feasible ones, which should be investigated in future works.

\section{Conclusion and outlook}\label{Summary}
We have introduced an inverse design framework that determines the microscopic structure of quantum wires based on magneto-conductance patterns. We have first demonstrated that our framework accurately reconstructs the defect from a measured magneto-conductance fluctuation. Next, we proved that our framework could automatically design onsite potentials for almost arbitrarily complicated magneto-conductance patterns. Finally, we demonstrated a technique for designing experimentally friendly structures by adding a regularization term to the objective function. 

Our results suggest several possible applications. For example, the identification of the defect using the observed magneto-conductance fluctuation can be applied to non-destructive measurements. In addition, the ability to design magneto-conductance fluctuations with specific characteristics can lead to the automated design of a magnetic sensor that exhibits a desired response. Furthermore, leveraging the capability to control the onsite potentials with an electric field could lead to programmable devices in which the response itself can be tuned on-demand by changing the electric field.

Further investigation is required to improve our framework. Regarding defect identification, increasing the number of leads could facilitate the detection of more complex defects. In the current setup with only two leads, the defect position along the lead direction could not be identified due to translational symmetry; however, this issue can be mitigated by attaching leads in a different direction. Additional extensions would lead to detecting multiple defects and structural modifications. For designing magneto-conductance fluctuations, optimizing parameters in the Hamiltonian beyond the on-site potential, such as the hopping term, could enable more flexible structural designs. Moreover, the development of regularization terms, as discussed in Sec. \ref{ResC}, will be crucial in future applications of our framework to experimental situations. {Additionally, it is important to address experimental uncertainties arising from impurities and other factors. One potential approach is to divide the process into two stages: first, constructing a tight-binding Hamiltonian through inverse analysis to reproduce the experimental results in the absence of a magnetic field, and then adjusting the on-site potential to replicate the desired magneto-conductance.}

In conclusion, automatic differentiation allows for the design of diverse physical properties in quantum nanomaterials. Our inverse design framework offers a powerful tool for exploring the relationship between microscopic structures and conductance patterns in quantum wires. Our strategy is advantageous over conventional ML methods in that it requires no training data, and it can be directly applied to a new material structure. In most of previous research, the conception of electronic devices has often relied on researchers' experience and intuition. The automation of system design, aiming to achieve the target quantum functionality through an inverse design approach, would facilitate the uncovering of novel functional quantum nanomaterials.

\section{Acknowledgement}
We are grateful to Shunsuke Daimon for fruitful discussions. Y.H. wishes to thank Yuya Hirata for providing valuable comments on the manuscript. This work was supported by the Center of Innovations for Sustainable Quantum AI (JST Grant No. JPMJPF2221), CREST (Nos. JPMJCR20C1, JPMJCR20T2) from JST, Japan; Grant-in-Aid for Scientific Research (S) (No. JP19H05600), Grant-in-Aid for Transformative Research Areas (No. JP22H05114), Grant-in-Aid for Early-Career Scientists (No. JP24K16985) from JSPS KAKENHI, Japan. 

\bibliography{hirasaki_manuscript}

\end{document}